\documentclass[twocolumn,aps,prl,showpacs,amsmath,superscriptaddress,longbibliography,notitlepage]{revtex4-2}
\usepackage{amssymb}
\usepackage{mathrsfs}
\usepackage{graphicx}
\usepackage{float}
\usepackage[caption=false]{subfig}
\usepackage[colorlinks,bookmarks=true,citecolor=blue,linkcolor=blue,urlcolor=blue]{hyperref}
\usepackage{verbatim}
\usepackage{epstopdf}
\usepackage[normalem]{ulem}
\usepackage{verbatim}
\usepackage{xcolor}
\usepackage{bm}
\renewcommand{\Re}{\operatorname{Re}}
\renewcommand{\Im}{\operatorname{Im}}

\begin{document}
\title{Tailoring Bound State Geometry in High-Dimensional Non-Hermitian Systems}
\author{Ao Yang}
\thanks{These two authors contributed equally}
\affiliation{Beijing National Laboratory for Condensed Matter Physics, and Institute of Physics, Chinese Academy of Sciences, Beijing 100190, China}
\affiliation{University of Chinese Academy of Sciences, Beijing 100049, China}

\author{Zixi Fang}
\thanks{These two authors contributed equally}
\affiliation{Beijing National Laboratory for Condensed Matter Physics, and Institute of Physics, Chinese Academy of Sciences, Beijing 100190, China}
\affiliation{University of Chinese Academy of Sciences, Beijing 100049, China}

\author{Kai Zhang}
\email{phykai@umich.edu}
\affiliation{Department of Physics, University of Michigan Ann Arbor, Ann Arbor, Michigan, 48109, United States}

\author{Chen Fang}
\email{cfang@iphy.ac.cn}
\affiliation{Beijing National Laboratory for Condensed Matter Physics, and Institute of Physics, Chinese Academy of Sciences, Beijing 100190, China}
\affiliation{Songshan Lake Materials Laboratory, Dongguan, Guangdong 523808, China}
\affiliation{Kavli Institute for Theoretical Sciences, Chinese Academy of Sciences, Beijing 100190, China}

\begin{abstract}
It is generally believed that the non-Hermitian effect (NHSE), due to its non-reciprocal nature, creates barriers for the appearance of impurity bound states. In this paper, we find that in two and higher dimensions, the presence of geometry-dependent skin effect eliminates this barrier such that even an infinitesimal impurity potential can confine bound states in this type of non-Hermitian systems. By examining bound states around Bloch saddle points, we find that non-Hermiticity can disrupt the isotropy of bound states, resulting in concave dumbbell-shaped bound states. Our work reveals a geometry transition of bound state between concavity and convexity in high-dimensional non-Hermitian systems, offering theoretical insights for the experimental manipulation of bound states.
\end{abstract}

\maketitle

\section*{Introduction}
The non-Hermitian Hamiltonian serves as an effective tool for describing systems that interact with environments~\cite{Rotter_2009,Diehl2011,Simon2015,Jean1992,Regensburger2012,Gao2015_Nature,FengLiang2017,Ganainy2018,Miri2019,YangLan2019,Kozii2024,ShenHT2018_PRL,FuLiang2020_PRL,SongF2019_PRL,Ashida2020}. Recently, non-Hermitian band systems have drawn much attention due to their intriguing phenomena that surpass the Bloch band framework~\cite{FuLiang2018_PRL,Sato2019_PRX}. A representative phenomenon is the non-Hermitian skin effect (NHSE)~\cite{Yao2018,Kunst2018_PRL,WangZhong2018,Murakami2019_PRL,ChingHua2019,LeeCH2019_PRL,LonghiPRR2019,Kai2020,Okuma2020_PRL,Slager2020PRL,Zhesen2020_aGBZ,Zhesen2020_SE,XuePeng2020,Ghatak2020,Thomale2020,LiLH2020_NC,Kawabata2020_Symplectic,Wanjura2020_NC,XueWT2021_PRB,LiLH2021_NC,Kai2022NC,ZhangDDS2022,Longhi2022PRL,YMHu2022}. In one dimension, the NHSE is characterized by a large number of eigenstates localized at the ends of an open chain, well understood in the generalized Bloch band framework~\cite{Yao2018,Murakami2019_PRL,Kai2020,Zhesen2020_aGBZ,Kawabata2020_Symplectic}. In higher dimensions, the NHSE exhibits more complexity due to the interplay between mode localization and boundary geometries. Particularly, the NHSE may disappear under certain geometry but reappear under others. This dimensionality-enriched phenomenon is referred to as the geometry-dependent skin effect (GDSE)~\cite{Kai2022NC,Kai2024edge,YangandKai2024,Wang2022NC,QYZhou2023NC,DingKun2023PRL,WanTuoSciB,QinYi2023PRA,GBJo2023arXiv}.

Impurities are fundamental in Hermitian systems and have been extensively studied for their broad applications. For example, magnetic impurities in metals induce phenomena such as the Kondo effect~\cite{Kondo2964}, while in s-wave superconductors, they manifest as Yu-Shiba-Rusinov bound states.~\cite{YU1965,Shiba1968,Rusinov1969}. Recently, the investigation of impurity states in non-Hermitian settings, especially their interplay with NHSE, has revealed various physical phenomena~\cite{Li2021scalefree,ChenShuScaleFree2023PRB}. A key aspect is that, NHSE creates barriers for the formation of impurity bound states due to its non-reciprocal nature~\cite{Hatano1996,Hatano1997_PRB}. Consequently, a finite impurity potential is necessary to induce a bound state when NHSE is present~\cite{Fang2023}. However, these phenomena have primarily been studied in 1D non-Hermitian systems, it is still unclear whether impurity states can exhibit different properties in higher dimensions. Additionally, NHSE presents unusual characteristics in higher dimensions~\cite{Kawabata2020, Kai2022NC,ContinuumBS2023PRL}, such as GDSE. The potential for impurity states to exhibit different behaviors in interaction with these emerging forms of NHSE in higher dimensions remains a significant and largely unexplored research gap. 

In this paper, we find that in the presence of GDSE, the impurity potential exhibits a zero threshold for the emergence of bound states. We establish an exact mapping between the bound state energy and the required impurity potential, demonstrating that even an infinitesimal impurity potential can confine bound states in a non-Hermitian system exhibiting GDSE. A key reason is that the GDSE ensures the presence of Bloch saddle points, which eliminate barriers to impurity-bound state formation.

In two and higher dimensions, the geometry of equal amplitude contours of wavefunction introduces a unique characteristic for non-Hermitian impurity-bound states. We determine the geometry of bound states using the mathematical concept of amoeba. In two dimensions, impurities can host anisotropic, concave bound states, in sharp contrast to the isotropic, convex bound states in Hermitian systems. Furthermore, we reveal a geometric transition from convexity to concavity in bound states by manipulating the impurity potential. This transition, characterized using our method, is observable in experimental setups, such as through local density of states patterns (See details in Supplementary Note. III).

\section{Result}
\subsection{A general theory of bound states in non-Hermitian systems}
\begin{figure}[b]
	\begin{center}
	\includegraphics[width=1\linewidth]{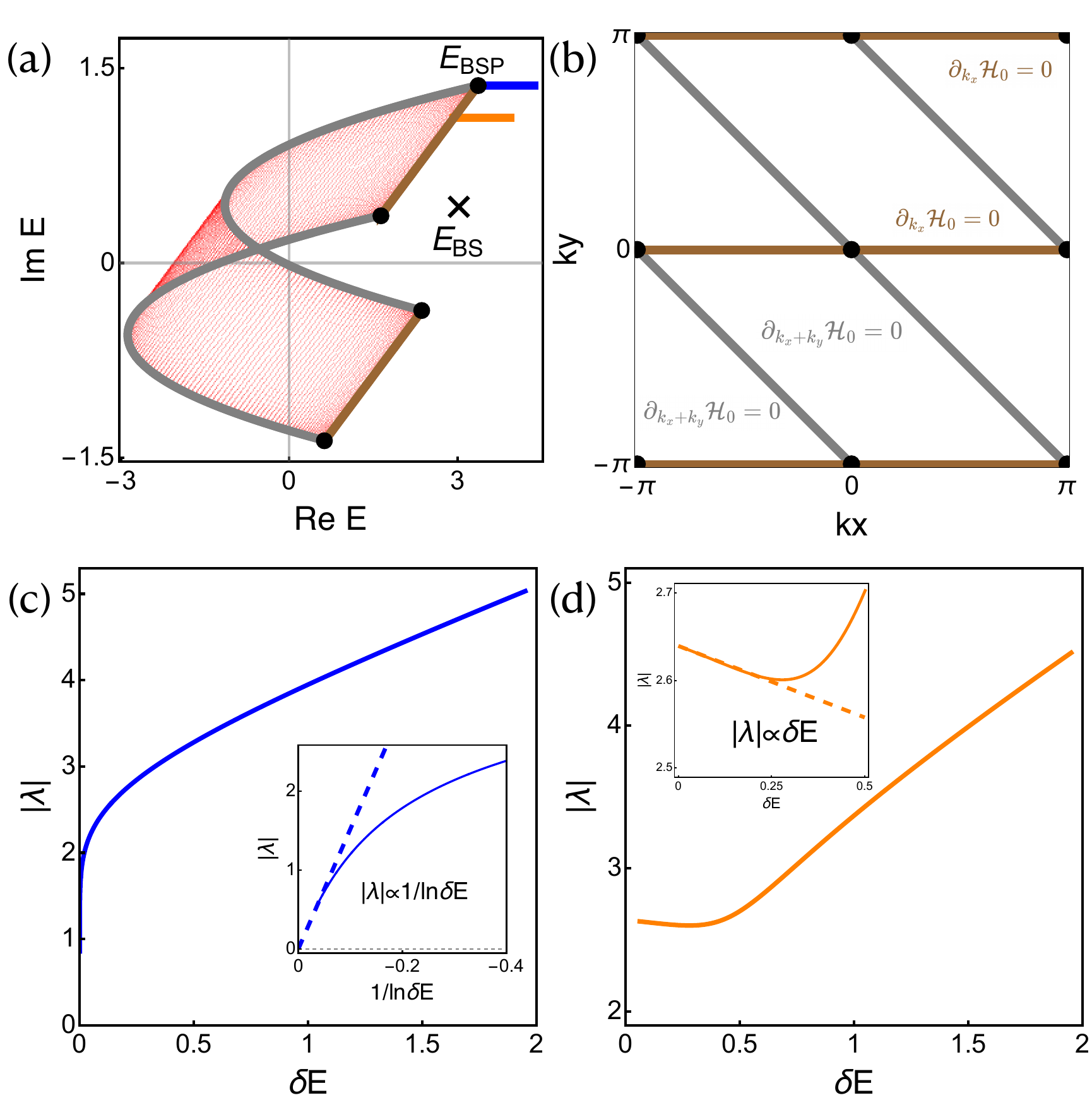}
	\par\end{center}
	\protect\caption{\label{fig:1}\textbf{Energy response of impurity strength.}
	(a) shows the PBC spectrum of the Hamiltonian $e^{i\pi/6}\cos{(k_x+k_y)} + e^{i\pi/3}\cos{k_x}+2\cos 2k_y$, with four black points denoting the energies at its BSPs. The black cross represents the bound state energy induced by the impurity.
	(b) illustrates the 1D Bloch saddle lines (BSLs) in the BZ, with brown lines representing $\partial_{k_y}\mathcal{H}_0(k_x,k_y)=0$ and gray lines for $\partial_{k_x+k_y}\mathcal{H}_0(k_x,k_y)=0$. The corresponding spectral lines $\mathcal{H}_0(k_x,k_y)$ are shown in the same color in panel (a). The four intersection points, i.e., high-symmetry $\textbf{k}$ points in the BZ, are the BSPs and correspond to the four vertices in the spectrum shown in (a). 
	(c) and (d) show the function $|\lambda(\delta E)|$, corresponding to the blue and orange trajectories in (a), respectively. Here, $\delta E$ is defined as $E - \mathcal{H}_0(0,0)$ in (c) and $E - \mathcal{H}_0(\pi/3,0)$ in (d). The insets in (c) and (d) show zoomed-in results as $|\lambda| \rightarrow 0$. } 
\end{figure}

We start from a general tight-binding Hamiltonian with finite range couplings in two dimensions, 
\qquad   \qquad  \qquad  \qquad   \qquad     
\begin{equation} \label{H0}
	\begin{split}
		H_0 &= \sum_{x,y}\sum_{s,l} t_{s,l} |x,y\rangle \langle x+s,y+l|  \\
		&= \sum_{k_x,k_y\in \text{BZ}} \mathcal{H}_0(k_x,k_y) |k_x,k_y\rangle\langle k_x,k_y|,
	\end{split}
\end{equation}
where $\mathcal{H}_0(k_x, k_y) = \sum_{l=-m,s=-n}^{l=M,s=N}t_{s,l}{(e^{ik_x})}^s {(e^{ik_y})}^l$, $(x, y)$ represents the position of lattice site, and $t_{s,l}$ indicates the hopping strength. 
The Bloch spectrum is formed by the eigenvalues of $\mathcal{H}_0(k_x,k_y)$ as $k_x$ and $k_y$ scan over the entire Brillouin zone (BZ), which we denote by $\sigma_{\text{PBC}}$ [red dots in Fig.~\ref{fig:1}(a)]. 
Note that though we focus on two-dimensional case here, the theory can be easily generalized to higher dimensions(see more detail in the Supplementary Note.I).

To generate an impurity bound state, we place a single impurity potential of strength $\lambda$ at the origin of the lattice, where the coordinate is set to $(x_0,y_0) = (0,0)$. 
The impurity potential takes the form 
\begin{equation} \label{latticemodel}
	V=\lambda \sum_{x,y} \delta(x,y) |x,y\rangle \langle x,y|.
\end{equation}
One can tune the impurity strength $\lambda$ such that the excited bound state has an energy $E_{\text{BS}}$ appearing beyond the region of $\sigma_{\text{PBC}}$ [the black cross in Fig.~\ref{fig:1}(a)]. 
Utilizing Green's function method, the wavefunction of this bound state can be analytically obtained as~\cite{wen2004quantum} 
\begin{equation}\label{Wavefunction}
	\psi_{E}(x,y)=\lambda \psi_{E}(0,0)G_0(E;x,y),
\end{equation}
where $G_0(E;x,y)=\langle x,y|1/(E-H_0)|0,0\rangle$ is the Green's function, $H_0$ is given by Eq.~(\ref{H0}), and $\psi_{E}(0,0)$ is determined by the wavefunction's normalization condition. 
Setting $x$ and $y$ to zero in Eq.~(\ref{Wavefunction}), the relationship between the bound state energy $E_{\text{BS}}$ and the required impurity strength $\lambda$ is established as
\begin{equation}\label{eq:BSRelation}
	\lambda^{-1}(E_{\text{BS}}) = G_0(E_{\text{BS}};0,0).
\end{equation}

Under PBC, the Green's function on the right-hand side of Eq.~(\ref{eq:BSRelation}) can be expanded under Bloch basis as an integral form, and thus the relationship becomes 
\begin{equation}\label{Impurity_and_BSenergy}
	\lambda^{-1}(E_{\text{BS}})= \int_{\text{BZ}} \frac{dk_x dk_y}{(2\pi)^2} \frac{1}{E_{\text{BS}}-\mathcal{H}_0(k_x,k_y)}.
\end{equation}
Typically, a state with energy within a continuum spectrum is expressed as a scattering state. Correspondingly, the energy of a bound state should lie outside the region of $\sigma_{\text{PBC}}$. 
The critical point, where the bound state energy merges with the PBC continuous spectrum, signifies a phase transition. This phase transition determines the minimum impurity strength required to create bound states. Consequently, we can define the set of minimum impurity strengths as
\begin{equation}\label{Impurity_Set}
	\Lambda=\left\{\lim_{E_{\text{BS}}\rightarrow E_b} \lambda(E_{\text{BS}}) \mid E_b\in \partial \sigma_{\text{PBC}} \right\},
\end{equation}
where $\partial \sigma_{\text{PBC}}$ represents the boundary of PBC continuum spectrum, and $E_b$ denotes a spectral boundary point. We define the impurity strength threshold $\lambda_0$ as the minimum absolute value $|\lambda|$ in the set $\Lambda$. The Bloch saddle points (BSPs), denoted as $(k_x^s, k_y^s)$, refer to the saddle points in the BZ where the relation holds: $\partial_{k_i}\mathcal{H}_0(k^s_x,k^s_y)=0$ for $i=x,y$. In the following, we demonstrate that zero threshold of impurity strength is ensured by the presence of BSPs in the Bloch spectrum $\sigma_{\text{PBC}}$. 

\subsection{The critical response to impurity potential near BSPs}
Here, we examine excitations around the BSP energy, assumed to be at the spectrum boundary $E_b$. 
The lattice Bloch Hamiltonian can be expanded at the BSP as $\mathcal{H}_0(q_x,q_y) = E_b + t ( q_x^2 + a\, q_y^2 + b\, q_x q_y)$, where $q_x$ and $q_y$ are deviations from the BSP momentum, and ${E_b, t, a, b}$ are expansion coefficients. 
The linear $q_x$ and $q_y$ terms are omitted since $E_b$ is a BSP. The $q_x q_y$ cross term can be eliminated through proper momentum basis rotation, when $\frac{|b|^2}{|a|^2}\leq \frac{(\sin(\arg a))^2}{\sin(\arg a-\arg b)\sin(\arg b)}$. 
Therefore, the expanded Hamiltonian around a BSP can be classified by the coefficients ${a, b}$. 
For demonstration, we utilize the following concrete lattice Hamiltonian:
\begin{equation}\label{eq: represented ham}
	\mathcal{H}_0(k_{x},k_{y})=\cos k_x+a\cos k_y+b \sin k_x \sin k_y.
\end{equation}
With a weak impurity potential, the excited bound state energy shifts slightly from the BSP energy $E_b=\mathcal{H}_0(0,0)$, i.e., $|\delta E| = |E_{\text{BS}}-E_b| \ll 1$. Substituting Eq.~(\ref{eq: represented ham}) into Eq.~(\ref{Impurity_and_BSenergy}), when $|b|,|\delta E|\ll 1$, the relationship between the impurity strength $\lambda$ and the bound state energy $E_\text{BS}$ becomes (see details in Supplementary Note. II): $ \lambda^{-1}(\delta E)=A\left( 5\ln2-\ln{B} - \ln{\delta E} \right)/2\pi$, where the parameters $A = \sqrt{(a-2 b^2)/(a^2-a b^2)}$ and $B = \sqrt{a (a+1)/(a-b^2)^2}$. 
Here, $\lambda^{-1}(\delta E)$ diverges asymptotically as $\ln{\delta E}$ when $\delta E \to 0$, which is expressed as: 
\begin{equation}\label{eq:BSP_Relation}
	\lambda^{-1}(\delta E) \propto \ln{\delta E}. 
\end{equation}
We emphasize that, as shown by Eq.(\ref{eq:BSP_Relation}), the bound state energy near the BSP is highly sensitive to the impurity potential, verified in Fig.~\ref{fig:1}(c). This contrasts sharply with the linear response observed near the regular spectrum boundary energy, depicted in Fig.~\ref{fig:1}(d). This sensitivity can be utilized to detect BSPs in higher-dimensional non-Hermitian systems\cite{QYZhou2023NC, DingKun2023PRL, WanTuoSciB}. As $\delta E$ approaches zero, the required impurity potential $\lambda$ also tends to zero, indicating a zero threshold at the BSPs, leading to the conclusion that BSPs ensure a zero threshold for impurity potential.

Numerical verification for the zero threshold at BSPs is illustrated in Fig.~\ref{fig:1}. As $E_{\text{BS}}$ approaches $E_{\text{BSP}}$ along the blue line in Fig.~\ref{fig:1}(a), the required impurity potential decreases to zero (Fig.~\ref{fig:1}(c)). Conversely, when $E_{\text{BS}}$ approaches a regular spectral boundary energy along the orange trajectory in Fig.~\ref{fig:1}(a), the impurity potential reaches a finite value (Fig.~\ref{fig:1}(d)). Dashed lines in Figs.~\ref{fig:1}(c) and (d) indicate the asymptotic behavior near the boundary.

A natural question arises: which systems ensure the existence of BSPs in higher dimensions? The answer lies in systems exhibiting GDSE. In GDSE, there are two special directions, $k_1$ and $k_2$. When boundary cuts are made along these directions, the open boundary eigenstates manifest as Bloch waves. By imposing open boundary conditions along $k_1$ and periodic boundary conditions along $k_2$, the $k_2$ momentum is conserved, treating the Hamiltonian as a 1D $k_1$-subsystem for a fixed $k_2$. With no skin effect in the $k_1$ direction, the energy spectrum forms an arc whose endpoints satisfy $\partial_{k_1} \mathcal{H}_0(k_x,k_y)=0$. As $k_2$ varies from $-\pi$ to $\pi$, these endpoints form two lines (brown lines in Fig.~\ref{fig:1}(b)). Similarly, two lines can be obtained for $k_2$ (gray lines in Fig.~\ref{fig:1}(b)). At their intersections, the BSP conditions $\partial_{k_i}\mathcal{H}_0(k^s_x,k^s_y)=0$ for $i=1,2$ are satisfied, corresponding to four BSPs within the BZ. An example with $\{k_1,k_2\}=\{k_x,k_x+k_y\}$ is shown in Fig.~\ref{fig:1}(b), where the intersections are marked by black dots.

\begin{figure*}
	\begin{centering}
	\includegraphics[width=1\linewidth]{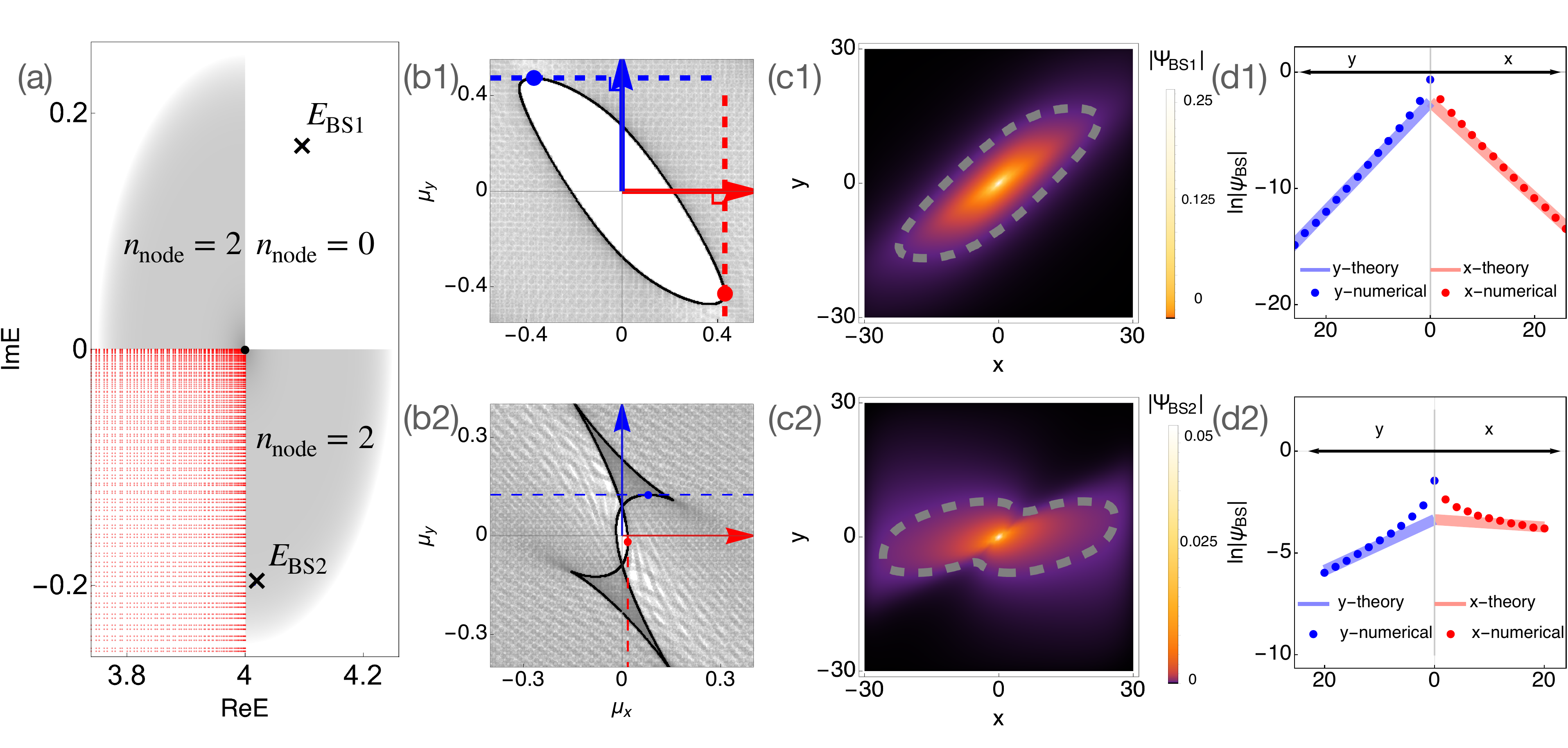}
	\par\end{centering}
	\protect\caption{\label{fig:2} \textbf{The relation between bound state's geometry and amoeba's contour. }
Parameters $\{t_{1,1}, t_{-1,-1}, t_{1,0}, t_{-1,0}, t_{0,0} \}$ for Hamiltonian in Eq.(1) are set to be $\{2 ,2, i, i, -2i\}$. 
(a) The red points represent the Bloch spectrum near the Bloch saddle point $\mathcal{H}_0(0,0)$ . The two gray regions indicate the range for energy whose amoeba has two nodes ($n_{\text{node}}=2$), which results in a concave wavefunction. And the white region is the range where the amoeba has no node ($n_{\text{node}}=0$). The impurity strength is $\lambda = 2.27+2.23i$ for bound state with energy $E_{\text{BS1}}=\mathcal{H}_0(0,0)+0.2\exp(i \frac{\pi}{4})$ and $ \lambda=2.66+0.96i$ for $E_{\text{BS2}}=\mathcal{H}_0(0,0)+0.2\exp(-i \frac{19}{40}\pi)$.
(b1) and (b2) show the corresponding amoeba's contours for $E_{\text{BS1}}$ and $E_{\text{BS2}}$ respectively outlined by the black curves. The red (blue) dot denotes the point of tangency between the red (blue) dashed line and the amoeba's contour. The red (blue) dashed line is perpendicular to the red (blue) arrow. 
(c1) and (c2) depict the amplitude $|\psi|$of the bound states  for $E_{\text{BS1}}$ and $E_{\text{BS2}}$  respectively. The gray dashed line is the equal amplitude curve of $|\psi(x,y)|$.
(d1) and (d2) show a comparison of bound states between the simulated data (colored dots) and the theoretical predictions (colored line).
The red (blue) dots and line correspond to the x(y)-axis. The slope of red (blue) line is given by the red(blue) point in (b1) and (b2).
The results are obtained from simulations performed on a $30\times 30$ lattice. }
\end{figure*}

\subsection{Tailoring the geometry of bound states.}
According to Eq.~(\ref{Wavefunction}), the bound state wave function is determined by Green's function. The Green's function can be expressed in an integral form with the Hamiltonian $H_0$ given by Eq.~(\ref{H0}): 
\begin{equation}\label{GF}
	\begin{split}
		G_0(E_{\text{BS}};x,y) &= \oint_{ \mathbb{T}^2}\frac{dz_x dz_y}{(2\pi i)^2 } \frac{e^{x\ln z_x+y\ln z_y}}{z_x z_y(E - \mathcal{H}_0(z_x,z_y))}
	\end{split}.
\end{equation}
Here, we extend the real momentum $k$ to the complex value $\tilde{k}_j=k_j+i\mu_j$ and define $z_j=e^{ik_j}$ for $j=x,y$. 
Under PBC, the integration contour is the BZ ($|z_x| = |z_y| = 1$), a torus in $\mathbb{C}^2$ space, denoted as $\mathbb{T}^2$. 
To compute this double integral, we adopt a step-by-step integration strategy. Firstly, we evaluate the first integral using the residue theorem. Noteondly, for the second integral, as we are primarily concerned with the asymptotic behavior of the wave function far from the impurity ($|x|, |y| \gg 1$), we can use the saddle-point approximation to handle the second integral, resulting in:
\begin{equation}\label{BSWave}
	\psi_{E_{\text{BS}}}(x,y)\propto G_0(E_{\text{BS}};x,y)\propto  e^{{ i \tilde{\mathbf{k}}_s(\theta) \cdot \mathbf{r}}}.
\end{equation}
Here, $\mathbf{r} = (x, y)$ and the complex momentum vector $\mathbf{\tilde{k}}_s(\theta) = (\tilde{k}_{s,x}(\theta), \tilde{k}_{s,y}(\theta))$ is a saddle point of the exponent $x\ln z_x+y\ln z_y=x\tilde{k}_x+y\tilde{k}_y$ in Eq.~\ref{GF} , depending on the spatial direction $\theta = \arg(\mathbf{r})$. 
Eq.~(\ref{BSWave}) demonstrates that the bound state wave function exhibits exponential behavior characterized by the complex momentum $\tilde{\mathbf{k}}$ along a fixed direction $\theta$, resulting in anisotropy in space. 
Therefore, we define the characteristic localization $\bm{l}$ that satisfies the relation: $|\psi_{E_{\text{BS}}}(l_{x}, l_{y})|/|\psi_{E_{\text{BS}}}(0,0)| = e^{-1}$, which is further expressed as: 
\begin{equation}\label{Localization}
	\mu_x l_x+\mu_y l_y=1. 
\end{equation}
Here, ${(l_x,l_y)}=l\,(\cos{\theta},\sin{\theta})$ forms a closed loop as $\theta$ changes, which characterizes the localization behavior and describes the geometric shape of impurity bound states. 

For a fixed direction $\theta$, the complex momentum $\tilde{\mathbf{k}}$ is determined by solving specific constraints (see details in the Supplementary Note. I). 
The first constraint is the bulk characteristic equation, 
\begin{equation}\label{amoeba}
	\begin{split}
	f(E_\text{BS},\tilde{k}_x,\tilde{k}_y)	= \det{ [E_{\text{BS}} - \mathcal{H}_0(\tilde{k}_x,\tilde{k}_y)] }= 0. 
	\end{split}
\end{equation}
The set of imaginary parts $(\mu_x,\mu_y)$ of the complex momentum $\tilde{\mathbf{k}}$ that satisfy the characteristic equation in Eq.~(\ref{amoeba}) is termed amoeba, as represented by the gray regions in Figs.~\ref{fig:2}(b1) and (b2). 
Since $E_{\text{BS}}\notin \sigma_{\text{OBC}}$, the corresponding amoeba always features a central hole~\cite{Wang2022}, as shown by the blank region in Fig.~\ref{fig:2}(b1) and (b2). 
Moreover, by solving Eq.~(\ref{amoeba}), $\tilde{k}_x$ or $\tilde{k}_y$ can be expressed as a function of the other. 
By applying the saddle point approximation $\partial_{\tilde{k}_y}(x\tilde{k}_x+y\tilde{k}_y)=0$ to the exponential factor in Eq.~(\ref{BSWave}) and utilizing the implicit function theorem $\nabla_{\mathbf{\tilde{k}}}\mathcal{H}_0 \cdot d \mathbf{\tilde{k}}=0$, the second constraint can be derived as 
\begin{equation}\label{amoebacontour}
	y \, \partial_{\tilde{k}_x}\mathcal{H}_0 - x \, \partial_{\tilde{k}_y}\mathcal{H}_0 = 0. 
\end{equation}
For the impurity bound state, the first constraint in Eq.(\ref{amoeba}) links the solution domain of ($\mu_x, \mu_y$) to the mathematical term amoeba; the second constraint in Eq.(\ref{amoebacontour}), combined with Eq.(\ref{amoeba}), identifies several isolated points $\boldsymbol{\mu}_s(\theta)$ on the amoeba. 
By varying the spatial direction $\theta=\arg \mathbf{r}$, $\boldsymbol{\mu}_s(\theta)$ forms a closed loop, corresponding to the amoeba's contour\cite{Passare2005,Gelfand2013,Krasikov2023}. 
In Figs.\ref{fig:2}(b1) and (b2), the amoeba's contours are depicted by the black curves. 
The constraint in Eq.(\ref{amoebacontour}) is a homogeneous function of $x$ and $y$, depending solely on the spatial direction $\theta=\arctan(y/x)$. 
Consequently, the bound state wavefunction exhibits exponential localization away from the impurity site but is anisotropic in real space.
Furthermore, by applying implicit function theorem $\nabla_{\mathbf{\tilde{k}}}\mathcal{H}_0 \cdot d \tilde{\mathbf{k}}=0$ to Eq.~(\ref{amoebacontour}), it can be transformed into the form $ \textbf{r}\cdot d\tilde{\mathbf{k}} = 0$. Notably, this is a complex equation, and by taking its imaginary part, we obtain
\begin{equation}\label{tangent}
	\textbf{r}\cdot d\bm{\mu} = 0. 
\end{equation}
This formula indicates that the inverse localization length $\boldsymbol{\mu}(\theta)$ of bound states along each spatial direction $\mathbf{r}$ is determined by the value $\boldsymbol{\mu}(\theta)=(\mu_x, \mu_y)$ on the amoeba's contour, where the tangent direction is perpendicular to $\mathbf{r}$.

As a result, for a fixed direction $\mathbf{r}$, we can determine the inverse localization length $\boldsymbol{\mu}(\theta)$ using Eq.~\ref{amoeba} and Eq.~\ref{amoebacontour}. By varying the spatial direction $\theta=\arg \mathbf{r}$, we find that the set of $\boldsymbol{\mu}(\theta)$ forms a closed loop on the amoeba, corresponding to the amoeba's contour. Additionally, Eq.~\ref{tangent} indicates that the tangent direction of $\boldsymbol{\mu}(\theta)$ is perpendicular to $\mathbf{r}$.

By substituting the values of $(\mu_x, \mu_y)$ into Eq.~(\ref{Localization}), we can determine the geometric shape of the bound state. 
As shown in Fig.~\ref{fig:2}(a), the bound states with energies $E_{\text{BS1}}$ and $E_{\text{BS2}}$ are depicted in Figs.~\ref{fig:2}(c1) and (c2), respectively, with their corresponding amoebas in Figs.~\ref{fig:2}(b1) and (b2). To further investigate localization behaviors, we plot $\ln{|\psi_{E_\text{BS}}(x,0)|}$ and $\ln{|\psi_{E_\text{BS}}(0,y)|}$ for these bound states in Figs.~\ref{fig:2}(d1) and (d2). Our findings indicate that the decay rates along the $x$ and $y$ directions are determined by points on the amoeba's contours, marked by red and blue dots in Figs.~\ref{fig:2}(b1) and (b2). Numerical verifications in Figs. ~\ref{fig:2}(d1) and (d2) show that the slopes at these contour points, represented by red and blue lines, match the numerical bound state wavefunctions, as indicated by the red and blue dots.

We conclude that the amoeba's contour encodes the localization lengths of the bound state along each spatial direction. Therefore, the amoeba's contour's geometric properties inevitably affect the shape of the wave function. Furthermore, since the amoeba's contour is uniquely determined by the bulk Hamiltonian of the system, it establishes a connection between non-Hermiticity and bound state geometric features.

\subsection{Geometry transition of bound state under weak impurity analysis.}
Using perturbation analysis with a weak impurity potential, we demonstrate a unique geometry transition in higher-dimensional non-Hermitian systems (see Supplementary Note. IV for strong impurity case ).
As mentioned, non-Hermitian systems with GDSE ensure the existence of BSPs, allowing bound state excitation by an infinitesimal impurity potential. 
Thus, GDSE systems provide a platform to examine bound state geometry with weak impurity excitation. 
The bound state geometry can be tailored by the amoeba's contour, determined by the characterization equation $f(E_{\text{BS}}, k_x, k_y) = E_{\text{BS}} - \mathcal{H}_0(k_x, k_y)$. We focus on its expansion near the BSP energy $E{\text{BSP}}$, generally expressed as:
\begin{equation}
	f(E_{\text{BS}},k_x,k_y) = E_{\text{BS}}-E_{\text{BSP}} - k_x^2 - e^{i \theta} k_y^2. 
\end{equation}
Here, the linear term of $k_x$ and $k_y$ vanishes due to the BSP condition, and cross term $k_x k_y$ is omitted for simplicity. 
By applying the constraints in Eq.~(\ref{amoeba}) and Eq.~(\ref{amoebacontour}), we can derive an algebraic curve of order 8 that describes the amoeba's contour (see details in Supplementary Note. II). 
Based on Eq.~(\ref{Localization}), we ultimately obtain an algebraic curve that features the bound state geometry shape: 
\begin{align}\label{WavShape}
	&[l_x^2 \sin{\alpha} + l_y^2 \sin(\alpha -\theta)]^2 -4 [l_x^2 \cos{\alpha} + l_y^2 \cos (\alpha -\theta ) ] = 4,
\end{align}
where $\alpha=\arg(E_{\text{BS}}-E_{\text{BSP}})$. This curve describes the localization length of the wave function along different directions and determines the shape of the bound states. 

When $\theta=0$, the Hamiltonian reduces to Hermitian, and the geometry curve collapses into a circle, given by
\begin{align}
	&\sin{\alpha}^2(l_x^2  + l_y^2)^2 -4  \cos{\alpha}(l_x^2 + l_y^2 ) = 4. 
\end{align}
In the Hermitian limit, the shape of the bound state is always circular or elliptical due to scaling factors on $k_x$ or $k_y$. 
However, when $\theta \neq 0$, varying $\alpha$ causes a transition in the bound state shape from a regular convex curve (Fig.~\ref{fig:2}(b1)) to a concave, dumbbell-like curve (Fig.~\ref{fig:2}(b2)), a feature unique to higher-dimensional non-Hermitian systems. This corresponds to the amoeba's contour transition from a regular curve (Fig.~\ref{fig:2}(b1)) to an irregular curve with multiple singular nodes (Fig.~\ref{fig:2}(b2)). A curve is convex if it has positive or negative curvature throughout its path, while singular nodes, which always appear in pairs due to reciprocity symmetry in GDSE systems, occur where the curve intersects itself. Concave geometry of bound states occurs if and only if the phase $(\alpha - \theta)$ or $(\theta - \alpha)$ falls within $(0, \theta)$, as detailed in the Supplementary Note. II. For weak impurity excitation near BSPs, when $\theta < \alpha < 2\theta$ or $-\theta < \alpha < 0$ (gray region in Fig.~\ref{fig:2}(a)), the amoeba contour shows two nodes, resulting in a concave bound state wave function geometry. 


\section{Conclusion}
In summary, we investigate impurity-induced bound states in 2D non-Hermitian lattice systems. The geometry of bound states is precisely determined by the corresponding amoeba. The presence of BSPs eliminates the threshold for the formation of impurity-bound states. The resulting bound state wavefunctions around BSPs can exhibit concave and anisotropic shapes, in stark contrast to the convex and isotropic configurations typically observed in Hermitian systems. Furthermore, we unveil a geometric transition from convexity to concavity in bound states by manipulating the impurity potential.  Since GDSE ensures the existence of BSPs, even an infinitesimal impurity potential in such systems can generate bound states near the BSP energy, making them ideal platforms for studying weak excitations. These findings demonstrate how non-Hermitian properties significantly enrich the geometric configurations of bound states.

\section{Data availability}
Data sharing not applicable to this article as no datasets were generated or analysed during the current study.

\section{Code availability}
All the computational codes that were used to generate the figures presented in this study are available from the corresponding authors upon reasonable request.

\section{References}
\bibliography{Refs_MainText}

\section{Acknowledgements}
C.F. acknowledges funding support by the Chinese Academy of Sciences under grant number XDB33020000
, National Natural Science Foundation of China (NSFC) under grant number 12325404, 12188101, and National Key R\&D Program of China under grant number 2022YFA1403800, 2023YFA1406704.

\section{Author contributions}
C.F. conceived the work; A.Y. did the major part of the theoretical derivation and numerical calculation; Z.F. wrote and analyzed the curve equation of weak impurity; K. Z. proposed and analyzed the critical response; All authors discussed the results and participated in the writing of the manuscript.

\section{Competing interests}
The authors declare no competing interests.

\newpage 
\appendix
\setcounter{equation}{0}  
\setcounter{figure}{0}  
\renewcommand{\thefigure}{A\arabic{figure}}
\renewcommand{\theequation}{A\arabic{equation}}
\begin{widetext}

{\begin{center}
		{\bf \large Supplementary Material for ``Tailoring Bound State Geometry in High-Dimensional Non-Hermitian Systems'' }
\end{center}}

\section{Appendix A. Proof of relation between the bound state and contour of amoeba}
In this section, we shall present our theory on geometry of bound state with more technical details. First, we shall explicitly formulate the wavefuntion following the method in the maintext. Then we move on to the proof for the relationship between geometry of wavefunction and amoeba's contour, i.e., the inverse of localization length at a given direction is determined by the point in the amoeba whose normal line is parallel to that direction. After that, some technical details concerning the deformation of the integration contour are discussed. 
\subsection{Formation of Geometry of Bound State}
We begin with the generalization of Green function method. The wavefunction is determined by
\begin{equation}
    \begin{split}
        \psi_{E}(x,y)&=\lambda \psi_{E}(0,0)G_0(E;x,y) \\
        &= \lambda \psi_E(0) \oint_{z_x,z_y\in \mathbb{S}^1}\frac{dz_x dz_y}{(2\pi i)^2 } \frac{z_x^x z_y^y}{z_x z_y(E - \mathcal{H}_0(z_x,z_y))}
    \end{split}
\end{equation}
where $\mathcal{H}_0(z_x \! \equiv \! e^{ik_x}, z_y \! \equiv \! e^{i k_y})\! =\! \sum_{l=-m,s=-n}^{l=M,s=N}t_{s,l}z_x^s z_y^l$ is the Hamiltonian in bivariate Laurant polynomial form defined in the main text. Integrate with respect to $z_y$ using residue theorem (here we assume $x,y>0$, so we choose residue inside the unit circle), 
\begin{equation}
    \begin{split}
        \psi_{E}(x,y) = \lambda \psi_E(0)&\oint_{z_x\in \mathbb{S}^1}\frac{dz_x}{2\pi i }\sum_{|z_y|<1}R(E,z_y(z_x))\times\\
		&e^{\mathrm{ln}z_x(x+m-1)+\mathrm{ln}z_y(y+n-1)}
    \end{split}
\end{equation}
Here $R(E,z_{y,i})=-1/t_{M,N}\prod_{j(\neq i)=1}^{N+n}(z_{y,j}-z_{y,i})$ is the residue of the function $z_x^m z_y^n (E-\mathcal{H}_0)$ at pole $z_{y,i}$, and the pole $z_{y,i}(z_x)$ is actually a function of $z_x$, i.e. , $z_{y,i}(z_x)$ are roots of $E-\mathcal{H}_0(z_x,z_y) = 0$ for $z_y$ given $z_x$. Since we only care about the asymptotic behavior at infinity, we can assume $x+m-1\approx x$, $y+n-1\approx y$ and evaluate above equation with saddle point approximation,  
\begin{equation}
    \begin{split}
        	\psi_E(x,y)=\lambda \psi_E(0)  \sum_{z_x,z_y}R_{E_{BS}}(z_x,z_y)z_x^x z_y^y
    \end{split}
\end{equation}
where $z_x,z_y$ is given by 
\begin{equation}
    \begin{split}
        	\frac{x}{z_x} + y \partial_{z_x}\mathrm{ln}z_y(z_x) =0
    \end{split}
\end{equation}
Denote $f(z_x,z_y) := E - \mathcal{H}_0(z_x,z_y) $ , and replace $\partial_{z_x}\mathrm{ln}z_y(z_x)$ in above equation with implicit function theorem, one can find
\begin{equation}
    \begin{split}
        \frac{x}{z_x}  - \frac{y}{z_y}  \frac{\partial_{z_x} f}{\partial_{z_y} f} =0
    \end{split}
\end{equation}
In next subsection, we shall see that this result is in accordance with the definition of amoeba's contour.

\subsection{Proof of the relation}
In this subsection, we shall prove that the saddle point determining the localization length of the bound state is given by points in the contour of amoeba. 

We shall begin with an introduction to the theory of amoeba contour ~\cite{Passare2005,Gelfand2013,Krasikov2023}. Consider a bivariate laurant form polynomial $f(z_x,z_y)= \sum_{t,s} a_{t,s} z_x^{t} z_y^s$, the contour of its amoeba is defined as 
\begin{equation}
	\begin{split}\label{eq:amoeba_contour}
		\mathcal{C}_f = \{(log|z_x|,log|z_y|)\in \mathbb{R}^2 | f(z_x , z_y) = 0 ,k z_x \partial_{z_x}f  - z_y \partial_{z_y} f=0 , k \in \mathbb{R} \cup \{\pm \infty\} \}
	\end{split}
\end{equation}
Here $k $ is a real parameter that encodes the slope of the normal line
to the contour of the amoeba. The boundary of amoeba is a subset of its contour $\mathcal{C}_f$. 

One can intuitively understand this definition by following considerations. Let a point $(z_{x0} ,z_{y0} )\equiv (e^{\mu_{x0} + i k_{x0}},e^{\mu_{y0} + i k_{y0}})$ sit at the boundary of the amoeba, then a neighbor point $(e^{\mu_{x0} + i k_{x0}+\delta \mu_x + i\delta k_x},e^{\mu_{y0} + i k_{y0}+\delta \mu_y + i\delta k_y})$ is mapped to 
\begin{equation}
	\begin{split}
		f(e^{\mu_{x0} + i k_{x0}+\delta \mu_x + i\delta k_x},e^{\mu_{y0} + i k_{y0}+\delta \mu_y + i\delta k_y}) = f(z_{x0} ,z_{y0} ) + (\delta \mu_x + i \delta k_x ) z_x \partial_{z_x} f + (\delta \mu_y + i \delta k_y ) z_y \partial_{z_y} f + \mathcal{O}(\delta \mu^2 , \delta k^2)
	\end{split}
\end{equation}
Giving a pair $(\delta \mu_x,\delta \mu_y)$ and treating $(\delta k_x,\delta k_y)$ as variables, one can always find a solution for $(\delta k_x,\delta k_y)$ such that $ (\delta \mu_x + i \delta k_x ) z_x \partial_{z_x} f + (\delta \mu_y + i \delta k_y ) z_y \partial_{z_y} f = 0$ if this equation is not degenerate. In other words, to find a neighbor point not in the amoeba, $(z_x \partial_{z_x} f,z_y \partial_{z_y} f)$ must be $\mathbb{R}-$ dependent, which gives the second relation in Supplementary Equation.~\ref{eq:amoeba_contour}.

In Supplementary Equation.~\ref{eq:amoeba_contour},  treating $f$ as giving an implicit function $z_x(z_y)$, then one get 
\begin{equation}
	\begin{split}\label{eq:saddle point}
		\frac{z_y}{z_x}\frac{dz_x}{dz_y} = -k =\frac{d \ln z_x}{d \ln z_y}= \frac{1}{2}(\frac{\partial \mu_x}{\partial \mu_y}- i \frac{\partial \mu_x}{\partial k_y}) + i \frac{1}{2}(\frac{\partial k_x}{\partial \mu_y}- i \frac{\partial k_x}{\partial k_y})
	\end{split}
\end{equation}

A few remarks shall be noted. First, the second term in the last equation naturally equals the first term by holomorphicity. Second, taking the imaginary part of the last equation above, one immediately gets $	\frac{\partial \mu_x}{\partial k_y} =0 $ . Taking the real part, one gets $k \frac{\partial \mu_y}{\partial \mu_x}  = -1 $, which implies $k$ is the slope of the normal line to the contour of the amoeba as we mentioned above.

\subsection{Some technical details concerning deforming integration contour}
In this subsection, we provide some technical details concerning the deformation of the integration contour in the calculation of the localization length.

Note that the statement that the localization length of the bound state is given by the saddle point of the integral is valid only when we can deform the contour of the integral to the saddle point without touching the poles of the integrand. In the following, we shall show that this is indeed the case since the poles are always discrete points in the complex plane for $z_y$.

Note that after integration of $z_x$ , we get (some constants are dropped for simplicity)
\begin{equation}
	\begin{split}
		\int_{z_y \in S^1} dz_y \sum_i \frac{z_{x,i}^s(z_y)z_y^t}{\prod_{j\neq i}(z_{x,i}(z_y)-z_{x,j}(z_y))}
	\end{split}
\end{equation}
As a function of $z_y$, $z_{x,i}(z_y)$ is analytical in the complex plane but for some discrete points. Indeed, $f(z_x,z_y)=E-\mathcal{H}(z_x,z_y)=0$ gives this implicit function and $\partial_{z_y}z_{x} = - \partial_{z_y} f / \partial_{z_x} f $ is not differentiable only when $\partial_{z_x}f = 0$ . Jointing this constraint with  $E-\mathcal{H}(z_x,z_y)=0$, one can only get finitely many points in the complex plane. 

The same argument holds for the numerator, since $z_{x,i} - z_{x,j}=0$ pose the other constraint and thus only leads to finite many poles.   

\subsection{Generalization to arbitray higher dimensions}
Our formulalism described above can be directly generalized to arbitray higher dimensions. Let's consider a $d-$dimensional lattice with bloch hamiltonian $h(\mathbf{z}), \mathbf{z}=(z_1,z_2,...,z_d), d \ge 2$. We shall focus on the asymptotic behavior of the wavefunction at infinity, i.e. , $\mathbf{r}=(r_1,r_2,...,r_d)=r(\alpha_1,\alpha_2,...,\alpha_d)$ with $\alpha_i$ the direction of the position satisfying $\sum_i \alpha_i^2 = 1$. The wavefunction is given by
\begin{align}
		&\int_{\times_{i=1}^dS^1} \frac{(\prod_{i=1}^{d}dz_i)}{(2\pi i)^d} \frac{z_1^{r_1} z_2^{r_2} ... z_d^{r_d}}{z_1 z_2 ... z_d (E-h(\mathbf{z}))} \\
        \propto&  \int_{\times_{i=1}^{d-1}S^1} (\prod_{i=1}^{d-1}dz_i) R(z_d,E) exp(r(\sum_{i=1}^{d} \alpha_i \ln z_i )), \label{eq:higher_dimensions}
\end{align}
where in Supplementary Equation.$\ref{eq:higher_dimensions}$ we've droped some unimportant constants and used the residue theorem to integrate out $z_d$ and denote $R(z_d,E)$ as the residue. We emphasize that in Supplementary Equation.$\ref{eq:higher_dimensions}$, the function $z_d(z_1,..,z_{d-1})$ is implicitly defined by $E-h(z_1,..,z_d)=0$. 

Denote the exponent as $g(z_1,..,z_{d-1})= \sum_{i=1}^d \alpha_i \ln z_i$ and it follows from saddle point approximation that for large $r$ the integral is dominated by the saddle point of $g(z_1,..,z_{d-1})$ given by 
\begin{equation}
    \begin{split}
       0&= \frac{\partial g}{\partial z_i} \\
       &= \frac{\alpha_i}{z_i}  + \frac{\alpha_d}{z_d} \partial_{z_i}  z_d \\
       &= \frac{\alpha_i}{z_i}  - \frac{\alpha_d}{z_d} \frac{\partial_{z_i} h}{\partial_{z_d} h} ,\quad \forall i\in \{1,2,...,d-1\}  \label{eq:higher_saddle}
    \end{split}
\end{equation}
where in the last equation we've used the implicit function theorem. Note that Supplementary Equation. \ref{eq:higher_saddle} is just equivalent to the definiton of contour of amoeba in higher dimensions,which is defined as the critical point of real logarithimic map $log|\cdot|$ ~\cite{Passare2005,Gelfand2013,Krasikov2023}.

\section{Appendix B. Discussion on weak impurity limit}
In this section, we shall discuss the weak impurity limit for GDSE case in more detail. We shall first perturbatively solve the contour of amoeba for such case and then we can get the condition for the argument of energy at which irregularity appears. After that, we will describe what can be expected from the bound state when one imposes a weak non-Hermicity to a Hermitian system and show the relationship between the argument of bound state energy and that of impurity. 

\subsection{GDSE Amoeba at weak impurity}
As shown in the main text, at weak impurity, the bound state takes energy near the Bloch Saddle Point and after a transforming of variables, we only need to consider the model $-2\cos k_x -2 e^{i \theta} \cos k_y$ , where a factor of 2 is added for the sake of simplicity. We point out here that this model has inversion symmetry $k_x \to -k_x, k_y \to -k_y$, thus the amoeba and its contour should also be symmetric with respect to inversion on x and y axis. Expand the 
Hamiltonian near the Bloch Saddle Point $(k_x,k_y) =(0,0)$, 

\begin{equation}\label{weak impurity original}
	\begin{split}
		h_0(\kappa_x,\kappa_y)=-2 - 2 e^{i \theta} +(k_x-i\mu_x )^2 + e^{i \theta} ( k_y-i\mu_y )^2 
	\end{split}
\end{equation}
with complex momentum $\kappa_{x(y)}=k_{x(y)}-i\mu_{x(y)}$.
By substituting this Hamiltonian into Equations 9 and 10 in the main text, we obtain equations for two complex number domains, that is, four equations for the real number domain. 
\begin{align}
    \label{eq:Ref}
    \Re\left( \delta E - (k_x-i\mu_x )^2 - e^{i \theta} ( k_y-i\mu_y )^2 \right)&=0\\
    \label{eq:Imf}
    \Im\left( \delta E - (k_x-i\mu_x )^2 - e^{i \theta} ( k_y-i\mu_y )^2 \right)&=0\\
    \label{eq:Redf}
    \Re\left( k \left(k_x-i \mu _x\right)-e^{i \theta } \left(k_y-i \mu _y\right) \right)&=0\\
    \label{eq:Imdf}
    \Im\left( k \left(k_x-i \mu _x\right)-e^{i \theta } \left(k_y-i \mu _y\right) \right)&=0
\end{align}
where $\delta E  = E -(-2 - 2 e^{i \theta})$. 
We can eliminate the variables $k_x$ and $k_y$ by solving Supplementary Equation.~\ref{eq:Redf} and Supplementary Equation.~\ref{eq:Imdf}, and substitute $k_x$ and $k_y$ into Supplementary Equation.~\ref{eq:Ref} and Supplementary Equation.~\ref{eq:Imf}. And because Supplementary Equation.~\ref{eq:Ref} and Supplementary Equation.~\ref{eq:Imf} have the same k for each $\mu_x,\mu_y$. Then we can use the resultant to find the final result $f_{\delta E}(\mu_x,\mu_y)=0$.
$f_{\delta E}(\mu_x,\mu_y)=0$ is a algebraic curve with parameters in $(|\delta E|, arg(\delta E))$. Further notice $f_{\delta E}(\mu_x,\mu_y)$ is homogeneous with respect to $(\mu_x ,\mu_y ,|\delta E|^\frac{1}{2})$ (which originates from Supplementary Equation. \ref{weak impurity original}) and one can perform the variable change $(\mu_x , \mu_y)\to (\mu'_x |\delta E|^\frac{1}{2}, \mu'_y|\delta E|^\frac{1}{2})$ to eliminate $|\delta E|$. Here we list the result (denote $\alpha := arg(\delta E)$ and we dropped superscript for simplicity)

\begin{align}\label{weak impurity contour}
		f(\mu_x,\mu_y) &=R(f_1(\mu_x,\mu_y,k),f_2(\mu_x,\mu_y,k),k)\\
            f_1(\mu_x,\mu_y,k)&=-k^2 \cos (\alpha )+k^2 \csc ^2(\theta ) \mu _x^2 \left(\cos (2 \theta )+k^2 \cos (\theta )\right)\nonumber\\
                        &-2 k \csc ^2(\theta )
   \mu _x \mu _y \left(\cos (\theta )+k^2\right)+\csc ^2(\theta ) \mu _y^2 \left(k^2 \cos (\theta )+1\right) \nonumber\\
            f_2(\mu_x,\mu_y,k)&=-k \csc (\theta ) \mu _x^2 \left(2 \cos (\theta )+k^2\right)+k \left(\sin (\alpha )+\csc (\theta ) \mu _y^2\right)+2
   \csc (\theta ) \mu _x \mu _y \nonumber
\end{align}
Here $f(\mu_x,\mu_y)$ is the equation of amoeba's contour, and $f_1(\mu_x,\mu_y,k)$ and $f_2(\mu_x,\mu_y,k)$ are deformations of Supplementary Equation.~\ref{eq:Ref} and Supplementary Equation.~\ref{eq:Imf}, and $R(f(x),g(x),x)$ is the resultant of the polynomials $f(x)$ and $g(x)$ with respect to the variable x. 
Supplementary Fig \ref{fig:f2} plots $f(\mu_x, \mu_y)$ for different $arg(\delta E)$ with fixed $\theta = \frac{\pi}{3}$. Only when $\delta E$ takes value outside spectrum, i.e. $arg(\delta E) \notin (0,\theta)$, can a central hole be found. As described in the main text ,one may also observe that when $arg(\delta E) \in (\theta,2\theta)$ (symmetrically, $arg(\delta E) \in (0,-\theta)$) , irregularity takes place at $x (y)$ axis. In the following we shall make this observation valid.

\begin{figure}[b]
    \centering
    \includegraphics[width=0.7\linewidth]{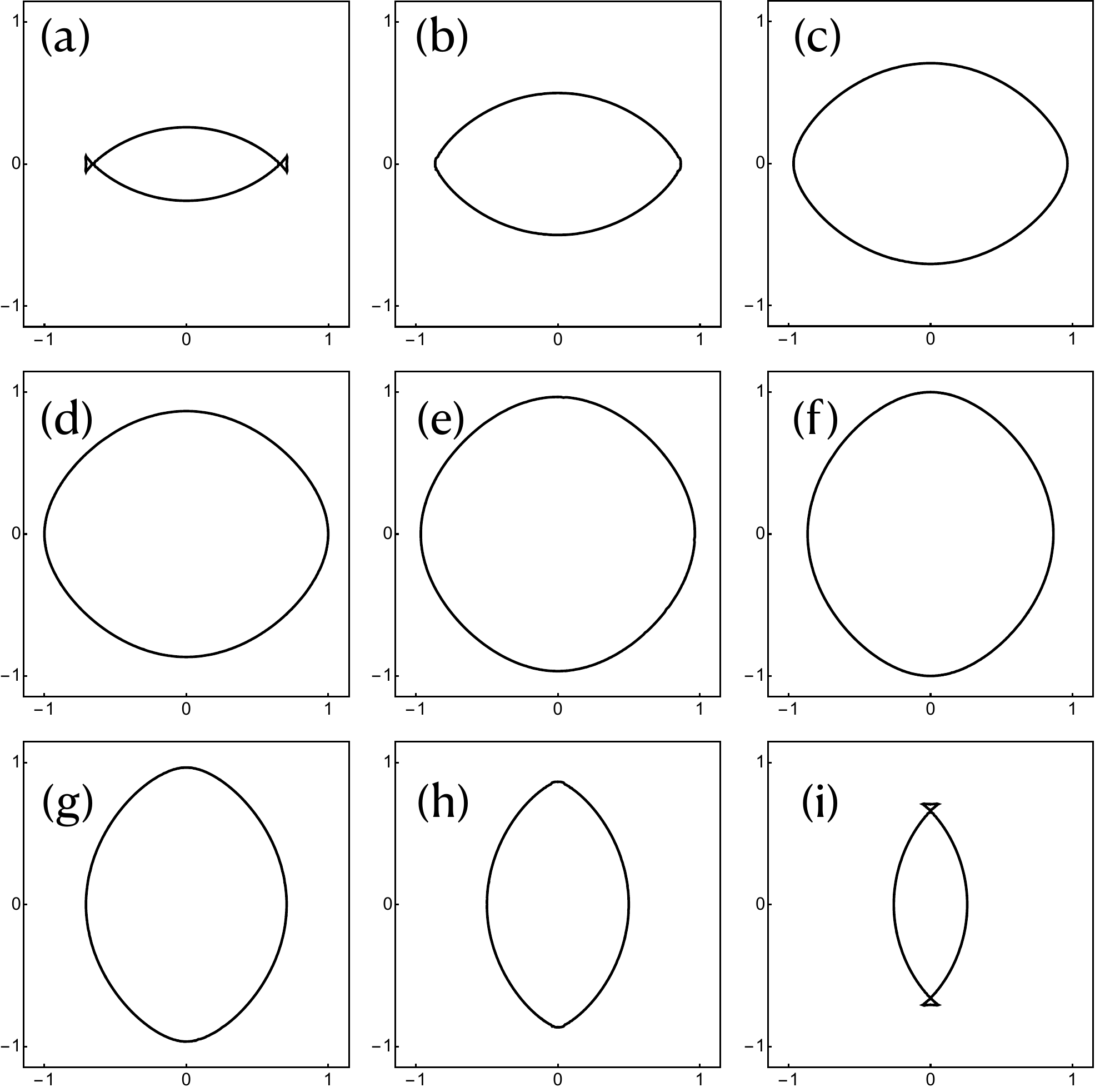}
    \caption{\textbf{amoeba's contour in Supplementary Equation.\ref{weak impurity contour} for different energy phase}. $f(\mu_x,\mu_y)=0$ in Supplementary Equation.\ref{weak impurity contour} for different $\alpha$. Fix $\theta = \frac{\pi}{3}$ (a)$\alpha = \frac{\pi}{2}$ ; (b)$\alpha = \frac{2\pi}{3}$; (c)$\alpha = \frac{5\pi}{6}$ ; (d)$\alpha = \pi$ ; (e) $\alpha = \frac{7\pi}{6}$ ; (f)$\alpha = \frac{4\pi}{3}$ ; (g)$\alpha = \frac{3\pi}{2}$ ; (h)$\alpha = \frac{5\pi}{3}$ ;(i)$\alpha = \frac{11\pi}{6}$ ;}
    \label{fig:f2}
\end{figure}

Let's take irregularity at x axis as an example and y axis can be treated similarly. Solve the equation $f(\mu_x,0)=0$,
\begin{equation} \label{weak impurity root}
    \begin{split}
        \mu_{x,1} = \sqrt{\frac{1-\cos(arg \delta E)}{2}} , \mu_{x,2} &= \sqrt{\frac{-1-\cos(arg(\delta E))}{2}},\mu_{x,3,4} = \sqrt{\frac{\sin(\theta)\sin(arg(\delta E)-\theta)}{2}}  \\
        \mu_{x,i} &= - \mu_{x,i-4},i= 5,6,7,8
    \end{split}
\end{equation}
Here we only need to deal with the first four roots since the remaining is just the inverse. Note that $\mu_{x,2}$ is always imaginary and $\mu_{x,1}$ is always real and regular. 
One can also check that $\mu_{x,3,4}$ appears as an irregular point, since it's also a solution for $\partial_{\mu_x}f =0 , \partial_{\mu_y}f=0$. Thus, to make $\mu_{x,3,4}$ real, one need $\sin(arg(\delta E)-\theta)>0$ (we can always assume $\theta \in (0,\pi)$). Then, recall that to make such an irregular point not isolated in the real plane, its Hessian matrix of second derivatives should have both positive and negative eigenvalues, and this gives the other side of the constraint. The Hessian matrix for $\mu_{x,3,4}$ reads as $diag(64 \sin(\theta)\sin(\theta-arg(\delta E))\sin^2(2 \theta - arg(\delta E)),64 \sin(\theta)\sin^3(2 \theta - arg(\delta E))$ . Take both constraints and the proposition is proved.

\subsection{Equal amplitude of wave function}
Because amoeba's contour determines the localization length (determined by equation $x\mu_x+y\mu_y=1$) of the wave function along a certain direction $y/x$, it is able to get the corresponding curves of the localization length of the wave function along different directions through amoeba's contour Supplementary Equation~\ref{weak impurity contour}, which is also the curve of the equal amplitude of the wave function.
\begin{equation}
    2 \left(\sin (\alpha ) x^2+\sin (\alpha -\theta ) y^2 \right)^2-8 \left(\cos (\alpha) x^2 +\cos (\alpha -\theta ) y^2 +1\right)=0
\end{equation}
This curve is deemed concave if it exhibits negative and positive curvature along its path, so the transition point between a concave curve and a convex curve is the point at which zero curvature first appears in the path. Using the Hessian matrix to calculate the curvature of the curve, we can find that zero curvature exists in the region with $0<|\alpha-\theta|<\theta$.

\subsection{Effect of weak non-Hermicity}
Let's first recast some facts on wavefunction in the Hermitian lattice case in the language of amoeba. In Supplementary Equation. \ref{weak impurity contour} , let $\theta =0$ and one get 
\begin{equation}
    \begin{split}
        f_{\theta =0}(\mu_x,\mu_y) = g(\mu_x^2 + \mu_y^2)
    \end{split}
\end{equation}
The precise form of the function $g$ is not important. What matters here is that $\mu_x^2 + \mu_y^2$ appears in the function as a whole due to symmetry, which we can also get from eq.\ref{weak impurity original}. This implies that the contour should be a circle, whose radius can be read from eq\ref{weak impurity root} as $\mu_{x,1}$. Thus, the wavefunction here is isotropic with localization length $\frac{1}{|\delta E|^\frac{1}{2}\mu_{x,1}(arg(\delta E))}$ . 

A few corollaries are ready in hand. First, comparing with the a real $\delta  E$, a complex one with same strength tends to suppress the localization of wavefunction. Second, when x and y are not symmetric, i.e., the hoppings are different, the contour of amoeba shall take the form of an ellipse and so will the geometry of wavefunction. One may also note that the mix term $\mu_x,\mu_y$ shall rotate the orientation of this ellipse.

Next, we fix the bound state energy real ($arg \delta E = \pi$) and show the result when one poses the non-Hermicity to the lattice. In eq.\ref{weak impurity contour}, fix a direction by defining $\mu_y = k \mu_x ,r^2 = \mu_x^2 + \mu_y^2$ and treat $f(\frac{r}{\sqrt{1+k^2}},\frac{kr}{\sqrt{1+k^2}})$ as an implicit function of $r (\theta)$ with parameters of $k$, allowing one to expand it at $\theta =0$ 
\begin{equation}
    \begin{split}
        r(\theta) = -\frac{1}{8} k^2(\frac{1}{1+k^2})^2(2+k^2) \theta^2 + O(\theta^4)
    \end{split}
\end{equation}
Note that the odd powers of terms naturally vanish since it's an even function. The negativity of the coefficient of the second order implies that in this case non-Hermicity also suppresses the localization of the bound state. Further notice $r''(\theta)|_{\theta=0}$ as a function of k is monotone in $(0,\infty)$ and $(-\infty,0)$ and vanishes at minimal $k=0$, so such suppression is not isotropic and do not influence localization at x axis, a fact which can also be retained from Supplementary Equation. \ref{weak impurity root}  and understood intuitively since non-Hermicity is only imposed to y axis. 

\subsection{relationship of the argument of bound state energy and impurity}
In the main text, we established the relationship between bound state energy and impurity near bloch saddle point via evaluating the integral with model $\cos k_x + a \cos k_y+b \sin k_x$ ,i.e.
\begin{equation}
    \begin{split}
        \lambda^{-1} &= \frac{1}{4\pi^2}\int_{BZ} \frac{dk_x dk_y}{1+1 a + \delta E -  \cos k_x - a \cos k_y-b\sin k_x} \\
        &=\frac{2 K\left(\frac{4 a\sqrt{1+b^2}}{(\delta E +1+\sqrt{1+b^2}) (\delta E +2a+1-\sqrt{1+b^2})}\right)}{\pi  \sqrt{(\delta E +1+\sqrt{1+b^2}) (\delta E +2a+1-\sqrt{1+b^2})}}
    \end{split}
\end{equation}
Note that $K(x)$, the Complete elliptic integral of the first kind, has asymptotic expansion near 1 as $K(1-x)\sim -\frac{ln(x)}{2}$,and one can get 
\begin{equation}
    \begin{split}
        \delta E \sim \frac{4 a e^{-\frac{\sqrt{a}\pi}{\lambda}}}{1+a}
    \end{split}
\end{equation}
Thus, the argument of bound state energy is linear with respect to the argument of impurity, i.e.
\begin{equation}
    \begin{split}
        arg(\delta E) \sim arg(\frac{a}{1+a}) - |\frac{\sqrt{a}}{\lambda}| (\frac{1}{2}arg(a)-arg(\lambda)) 
    \end{split}
\end{equation}
\begin{figure}[t]
    \begin{center}
        \includegraphics[width=0.7\linewidth]{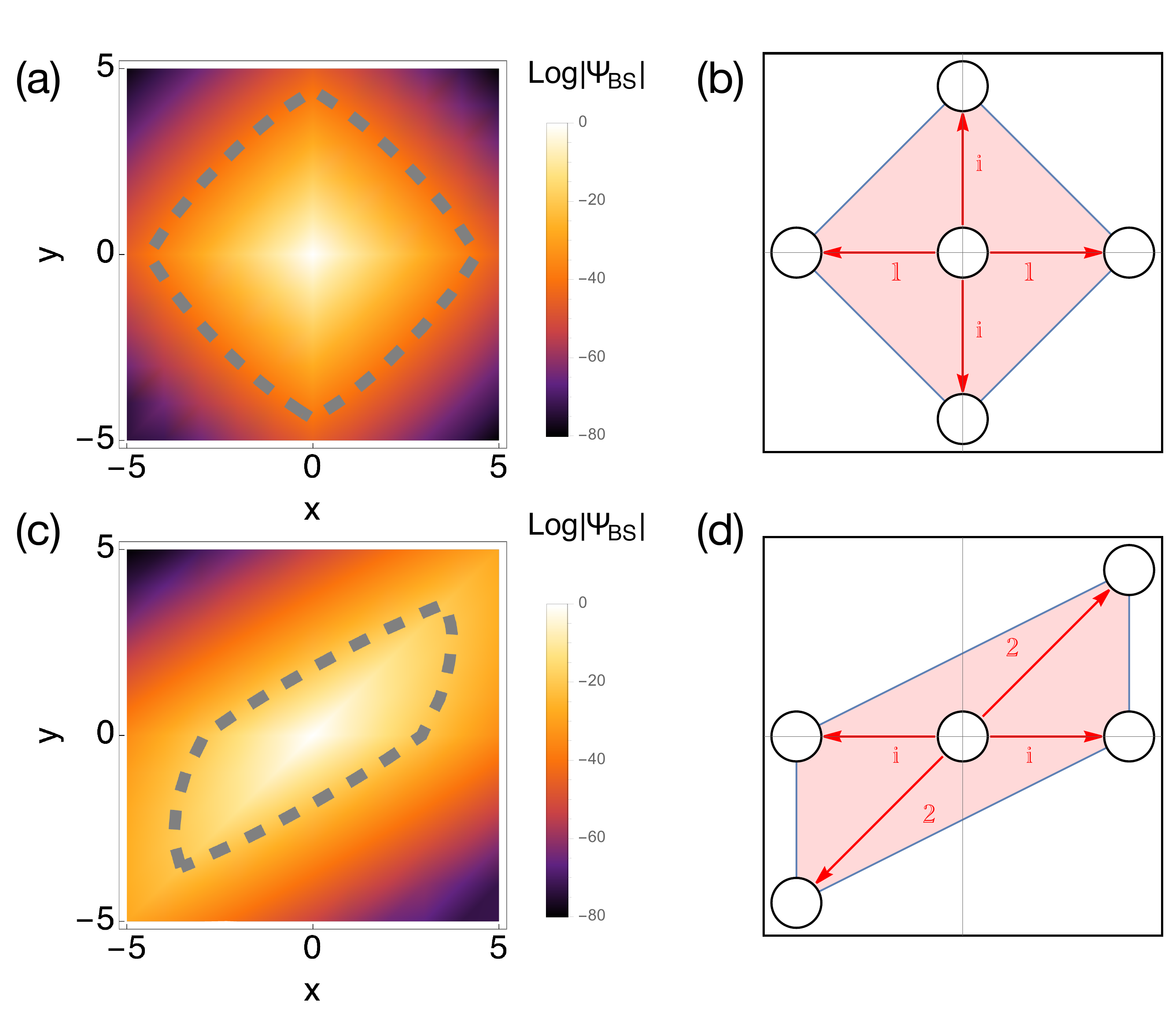}
        \par\end{center}
    \protect\caption{\label{fig:3} \textbf{The relation between bound state's shape and Newton polytope.} 
        Parameters $\{t_{1,1}, t_{1,0}, t_{-1,0}, t_{0, -1}, t_{0,1},t_{-1,-1} \}$ for Hamiltonian  are chosen to be $\{0 ,1, 1, i, i,0\}$ in (a) and (b), and $\{2, i, i, 0, 0, 2\}$ in (c) and (d). 
        Panels (a) and (c) depict amplitude of bound state wave function and the gray dashed lines represent the shape of bound state, while panels (b) and (d) show the corresponding Newton polytope. The red arrow in (b) and (d) represent the hopping direction and the red numbers denote the hopping strength in lattice model.
        The results are obtained from simulations performed on a $30\times30$ lattice, with data extracted from the central $11\times11$ sites with the impurity strength set to $\lambda = 50$.}
\end{figure}
\section{Appendix C. Local Density of States}
In this section, we will illustrate how the concavity and convexity of the wave function is reflected in LDOS by calculating LDOS.
We consider a non-Hermitian system $H_0$ with an impurity $V=\lambda|0,0\rangle\langle 0,0|$ at the origin. Through perturbation theory, we can know that when impurities are added, the Green's function of the system can be calculated as, 
\begin{equation}
    G(\omega; x, y)=g_0(\omega)+\frac{G_0(\omega; x, y ) G_0(\omega; -x, -y )}{1 / \lambda-g_0(\omega)}
\end{equation}
here $G(\omega;x,y)=\langle x,y|1/(\omega-H_0-V)|x,y\rangle$ represents the perturbed Green's function, $G_0(\omega;x,y)=\langle x,y|1/(\omega-H_0)|0,0\rangle$ represents the unperturbed Green's function with its entity at origin denoted as $g_0(\omega)=G_0(\omega;0,0)$.

From Eq.~3 in the text, we can know that $G_0(\omega;x,y)$ is proportional to the wave function of the bound state, and based on the previous discussion, we know that under the limit of weak impurity, the amoeba contour that determines the wave function has inversion symmetry about the origin. Therefore, our wave function should have inversion symmetry, so that our Green's function is proportional to the square of the wave function. As a result, LDOS will inherit the concavity and convexity of the wave function.

\section{Appendix D. Discussion on strong impurity case}

As the impurity strength increases, the shape of the bound state wave function progressively approximates a polygon and ultimately can be represented by a polytope known as the Newton Polytope, which is determined by the Bloch Hamiltonian.
The Newton Polytope of a polynomial is defined as the convex hull of the set of exponent vectors of monomials within it. 
Physically, the Newton Polytope is analogous to the convex hull of the hopping graph, which represents all the hopping terms in the Hamiltonian, as illustrated in Supplementary Fig~\ref{fig:3}(b) and (d).

As the impurity strength increases, the central hole of the amoeba expands and retreats to its spine, which is a dual representation of the system's Newton Polytope. Since the shape of the wave function is dictated by its localization length—a dual of the amoeba's contour, which encompasses the central hole, the wave function's shape resembles that of the Newton Polytope.

In Supplementary Fig.~\ref{fig:3}, we present the results for two examples of Hamiltonians. The first model incorporates the nearest hopping term, leading to a diamond-shaped Newton Polytope as illustrated in Supplementary Fig.~\ref{fig:3}(b). The corresponding wavefunction is plotted in Supplementary Fig.~\ref{fig:3}(a), exhibiting a diamond-like form. Conversely, the second model features a parallelogram-shaped Newton Polytope shown in Supplementary Fig.~\ref{fig:3}(d), accompanied by its parallelogram-like wave function depicted in Supplementary Fig.~\ref{fig:3}(c).

\end{widetext}

\end{document}